\documentclass[pre,floatfix,amsmath,twocolumn,showpacs]{revtex4}

\usepackage{calrsfs}
\usepackage{graphicx}

\newcommand{\nb}{\bar{n}}
\newcommand{\psib}{\bar{\psi}}
\newcommand{\kappab}{\bar{\kappa}}

\newcommand{\ie}{$i.\,e.$}
\newcommand{\eg}{$e.\,g.$}
\newcommand{\etal}{$et\,al.$}

\newcommand{\CALA}{{\mathcal{A}}}
\newcommand{\CALK}{{\mathcal{K}}}

\newcommand{\VECr}{{\boldsymbol{r}}}

\newcommand{\romB}{{\operatorname{B}}}
\newcommand{\romI}{{\operatorname{I}}}
\newcommand{\romK}{{\operatorname{K}}}

\newcommand{\romb}{{\operatorname{b}}}
\newcommand{\romc}{{\operatorname{c}}}
\newcommand{\romd}{{\operatorname{d}}}
\newcommand{\rome}{{\operatorname{e}}}
\newcommand{\romeq}{{\operatorname{eq}}}
\newcommand{\romf}{{\operatorname{f}}}
\newcommand{\romi}{{\operatorname{i}}}
\newcommand{\romlin}{{\operatorname{lin}}}
\newcommand{\romopt}{{\operatorname{opt}}}
\newcommand{\romr}{{\operatorname{r}}}
\newcommand{\roms}{{\operatorname{s}}}

\begin{document}

\title{The osmotic pressure of charged colloidal suspensions:\\
A unified approach to linearized Poisson-Boltzmann theory}

\author{Markus Deserno}
\affiliation{Department of Chemistry and Biochemistry, UCLA, USA}
\author{Hans-Hennig von Gr\"{u}nberg}
\affiliation{Fakult{\"a}t f{\"u}r Physik, Universit{\"a}t Konstanz, Germany}

\date{February 1, 2002}


\begin{abstract}
We study theoretically the osmotic pressure of a suspension of charged
objects (\eg, colloids, polyelectrolytes, clay platelets, etc.)
dialyzed against an electrolyte solution using the cell model and
linear Poisson-Boltzmann (PB) theory.  From the volume derivative of
the grand potential functional of linear theory we obtain two novel
expressions for the osmotic pressure in terms of the potential- or
ion-profiles, neither of which coincides with the expression known
from nonlinear PB theory, namely, the density of microions at the cell
boundary.  We show that the range of validity of linearization depends
strongly on the linearization point and proof that expansion about the
selfconsistently determined average potential is optimal in several
respects.  For instance, screening inside the suspension is
automatically described by the actual ionic strength, resulting in the
correct asymptotics at high colloid concentration.  Together with the
analytical solution of the linear PB equation for cell models of
arbitrary dimension and electrolyte composition explicit and very
general formulas for the osmotic pressure ensue.  A comparison with
nonlinear PB theory is provided.  Our analysis also shows that whether
or not linear theory predicts a phase separation depends crucially on
the precise definition of the pressure, showing that an improper
choice could predict an artificial phase separation in systems as
important as DNA in physiological salt solution.
\end{abstract}



\maketitle


\section{\label{sec:level1}Introduction}

In this paper we study the osmotic pressure of a suspension of charged
colloids or polyelectrolytes in osmotic equilibrium with an
electrolyte of given composition.  Examples of such systems abound in
our everyday life.  They occur as dispersion paints, viscosity
modifiers, flocculants, or superabsorbers, to name but a few
technological applications \cite{Hun94,DaJa94}.  They also play a
tremendous role in molecular biology, since virtually all proteins in
every living cell, as well as the DNA molecule itself, are charged
macromolecules dissolved in salty water \cite{LoBe01}.  A great deal
of experimental and theoretical research has been devoted to their
understanding, and several good textbooks
\cite{Hun94,DaJa94,Oos71,RuSa89,EvWe99,Rad01} and review articles
\cite{Kat71,BaJo96,HaLo00,Bel00} are available.

Arguably the most fundamental thing to know about these suspensions is
their equation of state, \ie, how the (osmotic) pressure depends on
other thermodynamic variables like macromolecular charge or
concentration.  Within the last hundred years several ingenious ways
have been conceived for treating this problem on varying levels of
sophistication.  In this article we will be concerned with
Poisson-Boltzmann (PB) theory in combination with a cell model
approximation for the macroion correlations, which we briefly revisit
in Sec.~\ref{sec:general}.  While this does not present the highest
level of accuracy or sophistication, it is probably the simplest and
up to today single most important starting point; it offers a
benchmark against which all other theories are compared.  Indeed, we
believe that modern improvements can only be fully appreciated once
one understands the successes and failures of the most fundamental
mean field theories.

Since the nonlinear PB equation can be solved analytically only in
very few cases \cite{And95}, its linearized version has always been an
important substitute.  However, the freedom to choose an expansion
point and its subsequent impact on the range of validity and accuracy
of the linearization has often gone unnoticed.  Moreover, the
computation of \emph{thermodynamic} properties from the ionic
\emph{profiles} computed in linear theory is often based on
expressions from the nonlinear PB theory or expansions thereof
\cite{PaGi72,RuBe81,StRi87,HaPo01}.  This procedure is by no means unique and
invariably entails internal inconsistencies.  Both these points make
it virtually impossible to conclude whether any failures of linearized
PB theory are real deficiencies or avoidable side-effects of a
non-optimal or inconsistent linearization.

In this paper we resolve these issues by giving a coherent
presentation of linearized PB theory which illuminates the subtle
interrelations between the Donnan equilibrium, microion screening,
linearization and the osmotic pressure.  In Sec.~\ref{sec:LPB} we
utilize the functional approach to PB theory
\cite{Lev39,BrRo73,ReRa90} and generalize its quadratic expansion
\cite{LoHa93}, arriving at a functional which yields the PB equation
linearized about the electrostatic potential value $\psib$.  This
expansion point will turn out to lie at the heart of all those
interrelations.  For general $\psib$ we then derive in
Sec.~\ref{sec:pressure} an analytical formula for the pressure in
terms of the ionic profiles.  Our novel expression replaces the famous
boundary density rule \cite{Mar55} from nonlinear PB theory, according
to which the osmotic pressure of the suspension is given by the value
of the microion density at the outer cell boundary.  That this does
not hold in \emph{linearized} theory may be considered as one of the
major results of the present work.

The most common choice of the linearization point $\psib$ -- namely,
the potential value in the salt reservoir -- suffers from several
drawbacks, as this value can be very different from the average
electrostatic potential $\langle\psi(\VECr)\rangle$ in the colloidal
suspension, and it seems more reasonable to selfconsistently linearize
about the latter \cite{RuBe81,TrHa96,TrHa97,GrRo01}.  We investigate
this choice in Sec.~\ref{sec:psibb} and will prove that it is indeed
optimal---in the sense that ($i$) charge screening rests on the actual
ionic strength, such that ($ii$) the crossover between counterion- and
salt-screening is naturally included and ($iii$) the limit of large
volume fraction is correctly reproduced.  In brief, all zeroth order
effects of the Donnan-equilibrium are already incorporated by the mere
choice of the linearization point, and the linearized equation now
describes higher order effects.  We will also see that ($iv$) all
other choices of the linearization point overestimate the Donnan
effect in lowest order, since they violate a rigorous inequality from
PB theory for the salt content in the colloidal suspension.

Based on this optimal linearization scheme we then derive explicit
analytical formulas for the osmotic pressure of suspensions of charged
mesoscopic objects in Sec.~\ref{sec:explicit}.  These formulas hold
for spherical, cylindrical and planar shapes and should thus be useful
in a broad variety of possible systems.  The subsequent
Section~\ref{sec:examples} is devoted to comparing their predictions
with the full nonlinear PB theory.

Within full PB theory the pressure is always positive.  Whether or not
this also holds in the linearized theory depends both on the choice of
the linearization point $\psib$ as well as on the precise definition
of the pressure itself.  We prove that the pressure is always positive
for symmetric electrolytes if one treats $\psib$ as an independent
variable.  If one does not, the pressure can become negative at low
volume fractions \cite{GrRo01,RoHa97,CaTr98,RoDi99,War00,DiBa01}.  The
implied liquid-gas coexistence -- not being present on the nonlinear
level -- is thus clearly an artifact.  As a striking example we show
that even a solution of DNA molecules under physiological conditions
would be predicted to phase separate at all relevant densities.


\section{General framework}\label{sec:general}

In this section we start by introducing the physical situation we wish
to describe -- namely, the Donnan equilibrium -- and its theoretical
description in terms of a cell model.  PB theory is founded on its
grand potential functional, and a brief derivation for the pressure
(leading to the boundary density rule) is presented.


\subsection{The Donnan equilibrium}

\begin{figure}
  \includegraphics[scale=0.5]{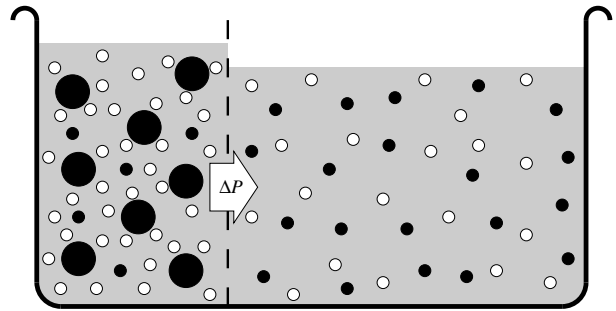}
  \caption{A solution of charged objects is in osmotic (Donnan-)
  equilibrium with a salt reservoir of given composition.  The
  membrane is permeable for small ions only and has to support an
  excess osmotic pressure $\Delta P$.  The mesoscopic objects are
  depicted as spherical (as appropriate for instance for
  ``conventional'' charged colloids or micelles), but our discussion
  will be more general and also apply to cylindrical entities (\eg\
  DNA, actin filaments, TMV viruses) or planar objects (\eg\ charged
  membranes, clay particles).}
  \label{FIG:1}
\end{figure}

We study a suspension of charged mesoscopic objects (henceforth simply
referred to as ``colloids'') dialyzed against a salt reservoir of
given composition, as illustrated in Fig.~\ref{FIG:1}.  This
situation is traditionally referred to as a ``Donnan membrane
equilibrium'' \cite{Don24,Ove56,TaLe98}.  Much of our discussion will
not depend on the shape of the colloids, and our final explicit
formulas will be valid for spherical, cylindrical and planar
geometries.  Even though the microions can traverse the membrane,
their average concentration differs between the salt reservoir and the
colloid compartment, since the latter is already occupied by the
counterions originating from the macroions, which cannot leave the
compartment due to the constraint of global electroneutrality.  This
imbalance in average densities generates an osmotic pressure
difference which the membrane has to sustain and which we wish to
calculate in the following.

For simplicity we assume the reservoir to be sufficiently large such
that its ionic strength remains unchanged after being brought in
contact with the macroion solution.  This assumption is not necessary,
but simplifies our discussion of general theoretical issues.  How it
can be avoided is demonstrated in Ref.~\cite{TaLe98}.  We also note
that we will not describe the solvent explicitly but rather replace it
by a continuum with relative dielectric constant $\varepsilon_\romr$.


\subsection{Cell-model and Poisson-Boltzmann theory}

Theoretical concepts like the cell-model or PB theory have long become
standard tools, so we will restrict ourselves to a brief description
and only provide the basic equations---essentially in order to
introduce our notation and terminology.  A recent and more detailed
exposition can be found in Ref.~ \cite{DeHo01}.

The cell model approximation attempts to reduce the complicated many
particle problem of interacting charged colloids and microions to an
effective one-colloid-problem.  It rests on the observation that at
not too low volume fractions the colloids -- due to mutual repulsion
-- arrange their positions such that each colloid has a region around
it which is void from other colloids and which looks rather similar
for different colloids.  In other words, the Wigner-Seitz cells around
two colloids are comparable in shape and volume.  One now assumes that
($i$) the total charge within each cell is exactly zero, ($ii$) all
cells have the same shape and ($iii$) for actual calculations one may
approximate this shape such that it matches the symmetry of the
colloid (for instance, spherical cells around spherical colloids).  If
the radius of the colloids is $r_0$, the cell radius $R$ is chosen
such that $\phi=(r_0/R)^d$ equals the volume fraction occupied by the
colloids.  Here, $d$ measures the ``dimensionality'' of the colloid in
the sense that $d=1$, $2$ and $3$ corresponds to planar, cylindrical
and spherical colloids.  If one ($iv$) neglects interactions between
different cells, the partition function finally factorizes in the
macroion coordinates, \ie, the thermodynamic potential of the whole
suspension is equal to the number of cells times the thermodynamic
potential of one cell.

Within each cell the small ions assume some inhomogeneous distribution
which arises from their interactions with themselves as well as with
the charged colloid.  Computing the corresponding partition function
is impracticable, since all ions are strongly correlated with each
other.  Poisson-Boltzmann theory is the mean-field route to
circumventing precisely this problem.  A very powerful way to
formulate it starts from a thermodynamic potential functional
belonging to the appropriate ensemble---which in our case is the
grand canonical one:
\begin{eqnarray}
  \beta \Omega & = & \int_V \romd^d r \;
  \bigg\{
    \frac{1}{2}\psi(\VECr)\Big(\rho_\romf(\VECr)+\sum_i v_i \, n_i(\VECr)\Big)
  \nonumber \\
  & & \qquad\qquad
    + \sum_i n_i(\VECr)\Big(\log\big(n_i(\VECr)\Lambda^d\big) - 1\Big)
  \nonumber \\
  & & \qquad\qquad
  - \sum_i \mu_i \, n_i(\VECr)\bigg\}.
  \label{eq:Omega}
\end{eqnarray}
The meaning of the symbols is as follows: $\beta \equiv 1/k_\romB T$
is the inverse thermal energy; $\psi(\VECr)$ is the local
electrostatic potential (made dimensionless by multiplication with
$\beta e$, where $e$ is the positive unit charge); the potential is
generated by both the fixed charge density $e \rho_\romf(\VECr)$
(located for instance on the colloid surface) as well as the
distributions $n_i(\VECr)$ of mobile ions of species $i$, which have a
signed valence $v_i$ and a chemical potential $\mu_i$; $\Lambda$ is
the thermal de Broglie wavelength of the small ions; and the region of
integration, $V$, is understood to be the space within one cell that
is actually accessible to the small ions.  The functional minimization
of (\ref{eq:Omega}) subject to the constraint of Poisson's equation
and charge conservation yields the set of Euler-Lagrange equations
\begin{equation}
  n_i(\VECr) \; = \;
  \Lambda^{-d}\,\rome^{\mu_i}\,\rome^{-v_i\psi(\VECr)} \; = \;
  n_{i,\romb}\,\rome^{-v_i\psi(\VECr)}.
  \label{eq:Boltzmann}
\end{equation}
The $n_{i,\romb}$ are the concentrations of ions of species $i$ in the
salt reservoir where the electrostatic potential has been assumed to
vanish.  Combining Eqn.~(\ref{eq:Boltzmann}) with Poisson's equation
results in the nonlinear Poisson-Boltzmann differential equation for
the potential $\psi(\VECr)$ in the region within the cell accessible
to the microions.  After introducing the Bjerrum length $\ell_\romB :=
\beta e^2/\varepsilon_\romr$, it is written as
\begin{eqnarray}
  \Delta \psi(\VECr) & = & -4\pi \, \ell_\romB \sum_i v_i \, n_i(\VECr)
  \nonumber \\
  & \stackrel{\text{(\ref{eq:Boltzmann})}}{=} &
   -\kappa^2 \,\frac{\sum_i v_i\,n_{i,\romb}\,
  \rome^{-v_i\psi(\VECr)}}{\sum_i v_i^2\,n_{i,\romb}},
  \label{eq:PB}
\end{eqnarray}
with the Debye screening constant of the reservoir defined as
\begin{equation}
  \kappa^2 \; := \; 4\pi\ell_\romB\sum_i v_i^2\,n_{i,\romb}.
\end{equation}
If we reinsert the solution of this equation back into the functional
(\ref{eq:Omega}) and use Eqn.~(\ref{eq:Boltzmann}), we obtain its
equilibrium (=minimum) value, which is the grand potential of
nonlinear PB theory:
\begin{eqnarray}
  \beta \Omega_\romeq & = &
  \frac{1}{2}\int_V \romd^d r \; \rho_\romf(\VECr)\psi(\VECr)
  \nonumber \\
  & & - \int_V \romd^d r \; \sum_i n_i(\VECr)\Big[1+\frac{1}{2}v_i\,\psi(\VECr)\Big].
  \label{eq:Omega_eq}
\end{eqnarray}

We want to close with the following remarks: The above variational
principle can be constructed starting from the PB equation (see for
instance Ref.~\cite{ReRa90} and references therein), but this need not
give a unique functional \cite{FoBr97} and (once $\Omega$ has been
identified with the grand potential) appears like an upside-down
explanation for the key initial equation (\ref{eq:Boltzmann}).
However, PB theory can be well justified by deriving the functional
(\ref{eq:Omega}) from the underlying \emph{Hamiltonian}.  For
instance, it can be obtained as the saddle point of the field
theoretic action \cite{PoZe88,CoDu92,NeOr00,BoAn00}, as a density
functional reformulation of the partition function combined with a
first order cumulant expansion of the correlation term \cite{LoHa93},
or from the Gibbs-Bogoljubov inequality applied to a trial product
state \cite{DeHo01}. Those approaches also show that PB theory
provides an upper bound of the exact thermodynamic potential.


\subsection{The pressure in Poisson-Boltzmann theory}\label{ssec:PB_press}

One advantage of the thermodynamic functional approach to PB theory is
that it becomes immediately clear what the pressure is---in the
present ensemble the derivative of the grand
potential~(\ref{eq:Omega_eq}) with respect to the volume.  It proves
convenient to rewrite this in terms of the \emph{functional}
(\ref{eq:Omega}), which can be achieved as follows.  The variation
upon some change in volume can be decomposed into the ``orthogonal''
changes
\begin{equation}
  \delta\Omega
  \; = \;
  \frac{\partial\Omega}{\partial V}\bigg|_{n_i(\VECr)}\!\!\!\!\delta V
  + \int_V \romd^d r \sum_i \frac{\delta\Omega}{\delta n_i(\VECr)}\bigg|_V\delta n_i(\VECr),
\end{equation}
where the first term contains any explicit dependence on the volume
(at fixed ion profiles) and the second part is the implicit dependence
through the ion profiles (at fixed volume).  However, since the
equilibrium distributions make the grand potential functional
stationary with respect to variations of the density profile at fixed
cell geometry, the implicit terms vanish.  Hence, the pressure is just
the negative derivative of the grand potential \emph{functional} with
respect to the cell volume, \emph{evaluated at the equilibrium
profile}.  The derivative $\partial/\partial V$ is understood to imply
a movement of the outer neutral cell boundary, which shall be located
at $r=R$, and which only occurs in the boundaries of the volume
integral in Eqn.~(\ref{eq:Omega}). Note that rewriting the
electrostatic energy in terms of the densities gives a double
integral, and the product rule then cancels the prefactor $1/2$ in
front of the term describing the electrostatic energy.  Putting
everything together, one arrives at
\begin{eqnarray}
  \beta P & = &
  -\frac{\partial \, \beta\Omega_\romeq}{\partial V}
  \; = \;
  -\frac{\partial \, \beta\Omega}{\partial V}\bigg|_\romeq
  \label{eq:P_PB_general} \\
  & = &
  - \sum_i n_i(R) \bigg[
    v_i\psi(R) +
    \log\big(n_i(R)\Lambda^d\big)-1 -
    \mu_i\bigg]_\romeq
  \nonumber \\
  & = &
  \sum_i n_{i,\romb}\,\rome^{-v_i\psi(R)}
  \label{eq:P_PB_psi} \\
  & = &
  \sum_i n_i(R).
  \label{eq:P_PB_n}
\end{eqnarray}
This is the well known result \cite{Mar55} that within PB theory the
pressure is given by the sum of the ionic densities at the cell
boundary.  It actually holds beyond the mean-field approximation
\cite{WeJo82}, but this will not be our concern in the following.
Its generalization for more complicated cells can for instance be
found in Refs.~\cite{Bel00,TrHa97}.

Eqns.~(\ref{eq:P_PB_psi},\ref{eq:P_PB_n}) give the pressure acting
within the macroion compartment.  The \emph{excess} osmotic pressure
across the membrane is the difference between this pressure and the
pressure in the salt reservoir.  The latter is obtained with
comparative ease, since the electrolyte is homogeneous and hence the
route via a density functional is unnecessary.  On the same level of
approximation as above it is given by the van't Hoff equation
\begin{equation}
  \beta P_{\text{res}} \; = \; \sum_i n_{i,\romb}.
\end{equation}
This implies for the excess osmotic pressure across the membrane
\begin{equation}
  \beta \, \Delta P
  \; = \;
  \sum_i n_{i,\romb}\big(\rome^{-v_i\psi(R)}-1\big)
  \; \ge \;
  0.
  \label{eq:PB_Pge0}
\end{equation}
The last inequality follows from $\rome^x\ge 1+x$ and the fact that
the salt reservoir is neutral, \ie, $\sum_i v_i n_{i,\romb}=0$.
Hence, within PB theory the excess osmotic pressure is always
nonnegative.


\section{The grand potential in linearized theory}\label{sec:LPB}

When studying linearized PB theory we want to benefit from the same
thermodynamic coherence as in the nonlinear case.  We can achieve this
aim by likewise founding the key equations on a suitable grand
potential functional---a possibility that has previously been pointed
out by L{\"o}wen \etal\ \cite{LoHa93}.

First observe that a functional leading to a linearized version of the
PB equation necessarily is quadratic in the densities.  The only part
in Eqn.~(\ref{eq:Omega}) for which this is not already true are the
entropy terms $n_i(\VECr)[\log(n_i(\VECr)\Lambda^d) - 1]$.  Let us now
introduce a set of densities $\nb_i$ by ``Boltzmann-weighting'' the
reservoir densities $n_{i,\romb}$ with an as yet unspecified potential
value $\psib$:
\begin{equation}
  \nb_i \; := \; n_{i,\romb} \, \rome^{-v_i \psib}.
  \label{eq:nib}
\end{equation}
If we expand the entropic terms in Eqn.~(\ref{eq:Omega}) about the
points $\nb_i$ up to quadratic order we obtain
\begin{eqnarray}
  \beta\Omega_\romlin & = &
  \int_V \romd^d r \;
  \bigg\{
  \frac{1}{2}\psi(\VECr)\Big(\rho_\romf(\VECr)+\sum_i v_i n_i(\VECr)\Big)
  \nonumber \\
  & & \hspace{-5em}
    - \frac{1}{2}\sum_i \nb_i\,\Big[1+2\,(1+v_i\psib)\,\frac{n_i(\VECr)}{\nb_i} -
      \Big(\frac{n_i(\VECr)}{\bar{n}_i}\Big)^2\Big]\bigg\}.
  \label{eq:Omega_lin}
\end{eqnarray}
We will refer to this expression as the grand potential functional of
linearized PB theory, since its functional minimization (again, under
the constraint of Poisson's equation and charge neutrality) leads to
\begin{equation}
  n_i(\VECr)
  \; = \;
  n_{i,\romb} \, \rome^{-v_i\psib} \, \big[1-v_i(\psi(\VECr)-\psib)\big].
  \label{eq:lin_Boltzmann}
\end{equation}
This is the Boltzmann relation~(\ref{eq:Boltzmann}) linearized about
the potential value $\psib$.  We want to stress right from the
beginning that we leave the value of $\psib$ unspecified for the time
being.  Linearized PB theory is not unique, it is a one-parameter
family labelled by the expansion point.  The by far most common choice
found in the literature is $\psib=0$ (strictly speaking, the potential
in the salt reservoir), but this is not the only conceivable (let
alone optimal) possibility.  In Sec.~\ref{sec:psibb} we will come back
to this issue in greater detail.

Combination of Eqn.~(\ref{eq:lin_Boltzmann}) with Poisson's equation
yields the \emph{linearized Poisson-Boltzmann equation}
\begin{eqnarray}
  \Delta\psi(\VECr) & \stackrel{\text{(\ref{eq:lin_Boltzmann})}}{=} &
  -\kappa^2\bigg\{
    \frac{\sum_i v_i\,\nb_i}{\sum_i v_i^2\,n_{i,\romb}}
   -\frac{\sum_i v_i^2\,\nb_i}{\sum_i v_i^2\,n_{i,\romb}}\big(\psi(\VECr)-\psib\big)
  \bigg\}
  \nonumber \\
  & = &
  \kappab^2 \, \big(\psi(\VECr) - \psi_\romi\big),
  \label{eq:LPB}
\end{eqnarray}
with the renormalized screening constant $\kappab$ and the
inhomogeneous term $\psi_\romi$ of the differential equation defined
as
\begin{eqnarray}
   \kappab^2 & = & \kappa^2\,\frac{\sum_i v_i^2\,\nb_i}{\sum_i v_i^2\,n_{i,\romb}}
   \; = \;
   4\pi\ell_\romB \sum_i v_i^2\,\nb_i
   \label{eq:kappab} \\
   \text{and}\quad\psi_\romi & = & \psib + \frac{\sum_i v_i\,\nb_i}{\sum_i v_i^2\,\nb_i}.
   \label{eq:psii}
\end{eqnarray}
Note in particular that $\kappab$ appears as a screening constant
calculated with the Boltzmann-weighted densities, $\nb_i$; it is hence
different from the secreening constant in the salt reservoir.  For the
special case of a $v:v$ electrolyte this simplifies to
\begin{equation}
  \Delta\psi(\VECr)
  \; \stackrel{v:v}{=} \;
  \kappab^2\psi(\VECr) \, - \, \kappab^2\Big[\psib-\frac{1}{v}\tanh(v\psib)\Big]
\end{equation}
with
\begin{equation}
  \kappab^2
  \; \stackrel{v:v}{=} \;
  \kappa^2\cosh(v\psib)
  \; \ge \;
  \kappa^2.
  \label{eq:kappab_v:v}
\end{equation}
If we reinsert the solution of Eqn.~(\ref{eq:LPB}) into the functional
(\ref{eq:Omega_lin}) and use Eqn.~(\ref{eq:lin_Boltzmann}) we obtain
the equilibrium grand potential of linearized PB theory:
\begin{eqnarray}
  \beta \Omega_{\romlin,\romeq} & = &
  \frac{1}{2}\int_V \romd^d r \; \rho_\romf(\VECr)\psi(\VECr)
  - V \, \sum_i \nb_i\Big[1+\frac{1}{2}v_i\,\psib\Big]
  \nonumber \\
  & & 
  +\frac{V}{2}\,\sum_i v_i \, \nb_i\big(1+v_i\,\psib\big)\big\langle\psi(\VECr)-\psib\big\rangle.
  \label{eq:Omega_lin_eq}
\end{eqnarray}
The angular brackets $\langle\cdots\rangle$ denote the spatial average
$\frac{1}{V}\int_V \romd^d r \; \cdots$ over the part of the cell
volume accessible to the ions.


\section{The pressure in linearized theory}\label{sec:pressure}

As mentioned in the introduction, the computation of the pressure in
linearized PB theory has often been based on formulas originating from
nonlinear case or expansions thereof \cite{PaGi72,RuBe81,StRi87,HaPo01}.  For
instance, one could use the predictions for boundary potential or
density from linearized theory and insert them into formulas
(\ref{eq:P_PB_psi}) or (\ref{eq:P_PB_n}), respectively.  However,
although both formulas coincide on the nonlinear level, they yield
different results once the linearized equation is used to compute
$\psi(R)$ or $n_i(R)$, since Eqn.~(\ref{eq:Boltzmann}) no longer
holds.

Here we circumvent this source of inconsistency by avoiding any
recourse to results from nonlinear PB theory.  Analogously to
Sec.~\ref{ssec:PB_press} we base the pressure on the volume-derivative
of the grand potential of linearized PB theory,
Eqn.~(\ref{eq:Omega_lin_eq}).  This leads to a formula which gives the
pressure as a function of the ionic profiles and which replaces the
boundary density rule (\ref{eq:P_PB_n}).


\subsection{Relevant thermodynamic variables}

Before we differentiate the grand potential we would like to pause for
a moment and discuss the issue of relevant thermodynamic variables,
since the pressure-formula will turn out to depend on our choice for
them.

The grand potential of nonlinear PB theory depends on volume $V$,
temperature $T$, colloid charge $Q$, volume fraction $\phi$ and the
set of chemical potentials $\mu_i$.  In addition to these variables
the grand potential underlying \emph{linear} theory also depends on
the linearization point, $\psib$.  This proves a relevant issue since
$\psib$ itself may depend on volume, see the following
Sec.~\ref{sec:psibb}.  For reasons which will become clear later it
turns out to be extremely useful to consider $\psib$ as an independent
variable rather than insert the functional dependence $\psib(V)$ into
the potential and thereby eliminate $\psib$.

With this in mind, the thermodynamic definition of the pressure within
linearized PB theory becomes
\begin{equation}
  P_\romlin^{(1)} \; := \;
  -\left(\frac{\partial\Omega_{\romlin,\romeq}(V,T,\ldots,\psib)}%
              {\partial V}\right)_{T,\ldots,\psib},
  \label{eq:P_psi_1}
\end{equation}
\ie, the partial derivative of the potential (\ref{eq:Omega_lin_eq})
with respect to the volume, keeping all other variables fixed.  This
amounts to the following procedure: If one wishes to calculate the
pressure at a given volume, one first chooses the desired
linearization point $\psib(V)$ at this volume, but \emph{fixes} it
subsequently.  Then one measures the change in the grand potential
upon slightly changing the volume.

\begin{figure}
  \includegraphics[scale=0.42]{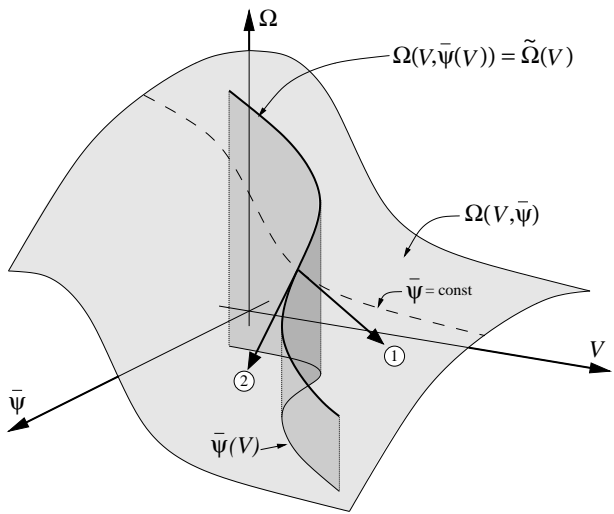}
  \caption{Relation between the two pressure definitions from
  Eqns.~(\ref{eq:P_psi_1}) and (\ref{eq:P_psi_2}) discussed in the
  text.  In the first scheme the derivative is taken in the direction
  of increasing $V$ along a coordinate line of constant $\psib$; in
  the second scheme the derivative is taken tangential to the
  submanifold $\Omega(V,\psib(V))$, which emerges from restricting the
  values $(V,\psib)$ to the ones given by the defining equation
  $\psib(V)$ of the linearization point.}\label{FIG:2}
\end{figure}

However, one could also argue that
$\tilde{\Omega}_{\romlin,\romeq}(V,\ldots) :=
\Omega_{\romlin,\romeq}(V,\ldots,\psib(V))$ is the desired grand potential.
In this case $\psib$ is not regarded as an independent variable, but
is removed from the description by substitution.  The pressure from
the derivative of this potential is then
\begin{eqnarray}
  P_\romlin^{(2)} & := &
  -\left(\frac{\partial\tilde{\Omega}_{\romlin,\romeq}(V,T,\ldots)}%
              {\partial V}\right)_{T,\ldots}
  \nonumber \\
  & = &
  -\left(\frac{\romd\Omega_{\romlin,\romeq}(V,T,\ldots,\psib(V))}{\romd V}\right)_{T,\ldots}
  \label{eq:P_psi_2} \\
  & = &
  P_\romlin^{(1)} \, - \,
  \left(\frac{\partial\Omega_{\romlin,\romeq}(V,T,\ldots,\psib)}%
             {\partial \psib}\right)_{V,T,\ldots} \!\!\! \frac{\romd\psib(V)}{\romd V}.
  \nonumber
\end{eqnarray}
This is the \emph{total} derivative of the grand potential
$\Omega_{\romlin,\romeq}(V,\ldots,\psib(V))$ with respect to volume,
and it differs from $P_\romlin^{(1)}$ by an additional term which
stems from the volume dependence of $\psib$.  The incorporation of the
constraint $\psib(V)$ amounts to the restriction of possible
$(V,\psib)$ values to a one-dimensional submanifold.  One hence
compares values of the grand potential at neighboring points on that
submanifold, along which one differentiates.
Fig.~\ref{FIG:2} gives a schematic illustration of the
difference between these two definitions.

After these considerations we can now proceed with our aim to derive a
formula which gives the pressure in terms of the solution of the
linear equation.


\subsection{The derivative of the functional}

The same line of reasoning that led to Eqn.~(\ref{eq:P_PB_general})
can be employed to rewrite the derivative of the equilibrium grand
potential as a derivative of the grand potential functional.  If one
again remembers that $\partial/\partial V$ just corresponds to a
movement of the outer neutral cell boundary, the pressure definition
(\ref{eq:P_psi_1}) and Eqn.~(\ref{eq:lin_Boltzmann}) give
\begin{eqnarray}
  \beta P_\romlin^{(1)}
  & = &
  \sum_i\Big[-v_i \, n_i(R) \psi(R) +
             \frac{\nb_i}{2}
  \nonumber \\
  & & \qquad
             +\;(1+v_i\psib)\,n_i(R) -
             \frac{1}{2}\frac{\big(n_i(R)\big)^2}{\nb_i}\Big]_\romeq
  \nonumber \\
  & = &
  \sum_i n_{i,\romb}\,\rome^{-v_i\psib} \, \Big[1-v_i\big(\psi(R)-\psib\big)
  \nonumber \\
  & & \qquad
             +\;\frac{1}{2}v_i^2\big(\psi(R)-\psib\big)^2\Big].
  \label{eq:P_LPB_1_psi} \\
  & = &
  \frac{1}{2} \sum_i n_i(R) \Big[\frac{n_i(R)}{\nb_i}+\frac{\nb_i}{n_i(R)}\Big].
  \label{eq:P_LPB_1_n}
\end{eqnarray}
Eqn.~(\ref{eq:P_LPB_1_psi}) replaces Eqn.~(\ref{eq:P_PB_psi}) and is
easily recognized as its \emph{quadratic} expansion about $\psib$.
Similarly, Eqn.~(\ref{eq:P_LPB_1_n}) replaces Eqn.~(\ref{eq:P_PB_n}).
Since $x+1/x \ge 2$, the expression~(\ref{eq:P_LPB_1_n}) is larger
than $\sum_i n_i(R)$, \ie, the expression that would follow if the
boundary density rule (\ref{eq:P_PB_n}) would also hold in the linear
case.

In the case of pressure definition (\ref{eq:P_psi_2}) we need the
explicit volume dependence of $\Omega_{\romlin,\romeq}$, which can
again be rewritten as the derivative of functional $\Omega_\romlin$,
Eqn.~(\ref{eq:Omega_lin}).  From Eqn.~(\ref{eq:nib}) follows
immediately $\partial \nb_i/\partial \psib = -v_i\nb_i$, and one
readily obtains using Eqns.~(\ref{eq:Omega_lin}) and
(\ref{eq:lin_Boltzmann})
\begin{equation}
  \frac{\partial \Omega_{\romlin,\romeq}}{\partial \psib}
  \; = \;
  \frac{\partial \Omega_\romlin}{\partial \psib}\bigg|_\romeq
  \; = \;
  \frac{1}{2}V\Big\langle\big(\psi(\VECr)-\psib\big)^2\Big\rangle\sum_i v_i^3\,\nb_i.
  \label{eq:dOmlindpsib}
\end{equation}
Combining this with Eqn.~(\ref{eq:P_psi_2}) results in an explicit
formula for $P_\romlin^{(2)}$:
\begin{equation}
  \beta P_\romlin^{(2)} \; = \;
  \beta P_\romlin^{(1)} \, - \, 
  \frac{1}{2}V\frac{\romd\psib}{\romd V}\Big\langle\big(\psi(\VECr)-\psib\big)^2\Big\rangle\sum_i v_i^3\,\nb_i.
  \label{eq:P_LPB_2}
\end{equation}
Note that unlike $P_\romlin^{(1)}$ this expression depends on the
whole potential distribution, and not just on the boundary potential
or the boundary densities.

Equations (\ref{eq:P_LPB_1_psi},\ref{eq:P_LPB_1_n}) and
(\ref{eq:P_LPB_2}) are a first key result of this paper.  They replace
the boundary density rule (\ref{eq:P_PB_n}) from nonlinear
PB theory, whose validity in the linear case has often falsely been
taken for granted.  In the second part of this article we will use
these equations to derive explicit formulas---once we have established
an optimal linearization point $\psib$, which will be the topic of the
next section.


\section{Optimal choice of $\psib$}\label{sec:psibb}

Up to now we have not specified the linearization point $\psib$;
rather, we have emphasized that its choice is largely at ones
disposition.  However, not all choices may be equally successful.  In
fact, the range of validity of linearization depends strongly on the
choice of $\psib$, since it is the \emph{difference} between the
potential and its linearization point that is required to be small and
not the potential itself.  This is particularly important in
concentrated suspensions where this difference can be small even when
the total potential is quite large.

In this section we identify an optimal linearization scheme---which
however first requires a clarification of what ``optimal'' is supposed
to mean.  It proves instructive to first study the obvious (but
futile) attempt to base optimality on a minimization of the grand
potential, as we discuss in Sec.~\ref{ssec:grand_pot_min}.  This sheds
some surprising light onto traditional linearization.  In
Sec.~\ref{ssec:psiopt} we will then discuss a scheme based on the
selfconsistently determined average potential and present its optimal
aspects in Sec.~\ref{ssec:optimal}, which even though this approach
has been used in the past \cite{RuBe81,TrHa96,TrHa97,GrRo01} have
largely gone unnoticed.


\subsection{Minimizing the grand potential?}\label{ssec:grand_pot_min}

One may try to obtain an optimal expansion point by looking for the
value of $\psib$ which minimizes the grand potential of linear theory.
Let us thus set the derivative of $\Omega_{\romlin,\romeq}$,
Eqn.~(\ref{eq:Omega_lin_eq}), with respect to $\psib$ to zero.  Using
Eqn.~(\ref{eq:dOmlindpsib}) and remembering that
$\langle(\psi(\VECr)-\psib)^2\rangle$ is nonzero unless the profile is
completely flat, we see that the value of $\psib$ at the extremum is
given by the solution of $\sum_i v_i^3n_{i,\romb}\,\rome^{-v_i\psib} =
0$.  The left hand side is a sum of strictly monotonically decreasing
functions.  If ions of both sign are present, the left hand side hence
falls monotonically from $+\infty$ to $-\infty$ and the equation has a
unique solution.  Unfortunately, however, this monotonic
\emph{decrease} of the derivative also implies that this solution
corresponds to a \emph{maximum} of the grand potential.  For a
symmetric $v:v$ electrolyte the solution is for instance given by
$\psib=0$.  The most widely used choice of the expansion point hence
gives the \emph{largest} grand potential of all possible linearization
schemes (this can for instance be verified in Fig.~1 of
Ref.~\cite{GrRo01}, which shows the thermodynamic potential for
various linearization schemes).

We hasten to remark that the above finding has to be put in the
correct perspective.  Minimization is only a meaningful venture if one
can be sure about the existence of a lower bound \footnote{Note that
for any \emph{given} value of $\psib$ the grand potential from
linearized PB theory is of course bounded below.}.  For
instance, the PB functional is bounded below by the exact
thermodynamic potential of the restricted primitive model.  Minimizing
the functional stems from the desire to get as close to this result as
is possible within a mean field description \cite{DeHo01}.  For the
parameter $\psib$ from linearized PB theory such a lower bound for the
functional cannot be constructed, hence minimization is meaningless.
And even if there were a bound, there would be no reason to approach
it---unless one knows that it is favorably related to the actual
thermodynamic potential.


\subsection{Expansion about the average potential}\label{ssec:psiopt}

Having seen that a minimization condition on $\Omega_{\romlin,\romeq}$
is not successful, we will approach the problem from a different
direction.  If we average Eqn.~(\ref{eq:lin_Boltzmann}) over the cell
volume and use Eqn.~(\ref{eq:nib}), we find
\begin{equation}
  \big\langle n_i(\VECr) \big\rangle
  \; = \; 
  \nb_i \, - \, v_i \, \nb_i \big\langle \psi(\VECr) - \psib \big\rangle.
  \label{eq:naverage}
\end{equation}
If we were to choose $\langle \psi(\VECr) - \psib \rangle=0$, the
second term would vanish and the expansion points for the densities,
$\nb_i$, would coincide with the averages $\langle n_i(\VECr)
\rangle$.  It clearly makes sense to expand about these average
values, since then the differences between actual value and expansion
point can be kept small \emph{throughout the cell}, \ie, in the whole
region in which linearization must work.  Hence, the choice
\begin{equation}
  \psib \; = \; \psib_\romopt \; := \; \langle \psi(\VECr) \rangle
  \label{eq:psibopt}
\end{equation}
is a particularly suitable one, which in anticipation of the results
from Sec.~\ref{ssec:optimal} we have labelled with the index ``opt''.

At first sight this particular choice may seem difficult to work with,
since the average $\langle\psi(\VECr)\rangle$ determines the value of
$\psib_\romopt$, but the potential $\psi(\VECr)$ to be averaged is in
turn the solution of the equation linearized about $\psib_\romopt$.
However, the necessary selfconsistency is readily fulfilled, since
$\langle \psi(\VECr) \rangle$ (and thus $\psib_\romopt$) does only
depend on the state point and not on the actual ionic profile: For
reasons of electroneutrality the total charge of all ions in the cell
must be the negative of the colloid charge, and so we have
\begin{equation}
  -\frac{Q}{V}
  \; = \;
  \sum_i v_i \langle n_i(\VECr) \rangle
  \; \stackrel{(\ref{eq:naverage},\ref{eq:psibopt})}{=} \;
  \sum_i v_i \, n_{i,\romb}\,\rome^{-v_i\psib_\romopt}.
  \label{eq:psibopt_determine}
\end{equation}
Satisfying this equation is necessary and sufficient for the validity
of Eqn.~(\ref{eq:psibopt}), so we can determine $\psib_\romopt$ also
from Eqn.~(\ref{eq:psibopt_determine}).  If the electrolyte contains
ions of both sign, the right hand side monotonically decreases from
$+\infty$ to $-\infty$ as a function of $\psib_\romopt$, so a solution
always exists and is unique.  For the special case of a $v:v$
electrolyte this equation reads
\begin{equation}
  \frac{Q/vV}{2n_\romb} \;\; \stackrel{v:v}{=} \;\; \sinh(v\psib_\romopt).
  \label{eq:psibopt_determine_v:v}
\end{equation}


\subsection{Optimality of $\psib_\romopt$}\label{ssec:optimal}

The linearization scheme from the above Sec.~\ref{ssec:psiopt} has
been put forward several times in the past \cite{RuBe81,TrHa96,TrHa97,GrRo01},
but various of its special properties have gone unnoticed.  In this
section we show, in which sense this scheme can be regarded as
optimal.

Since the choice of $\psib_\romopt$ implies that the expansion points
$\nb_i$ coincide with the average microion densities $\langle
n_i(\VECr)\rangle$, this translates to the corresponding screening
constant $\kappab_\romopt$ which -- as Eqn.~(\ref{eq:kappab}) shows --
is now also calculated with the average ionic strength within the
cell:
\begin{equation}
  \kappab_\romopt^2
  \; = \;
  4 \pi \ell_\romB \sum_i v_i^2\,\langle n_i(\VECr)\rangle.
\end{equation}
Due to the Donnan effect the latter is different from the one in the
reservoir, and it is more appropriate to have a linearization scheme
which derives its screening constant from the actual ion densities in
the macroion compartment.  In fact, describing the colloidal system in
an integral-equation approach together with the MSA closure yields an
effective pair potential between colloids that is a screened Coulomb
potential with the screening constant being equal to
$\kappab_\romopt$---see for instance Ref.~\cite{Bel00}.

Let us make the above remarks more explicit by specializing to a
symmetric $v:v$ salt of concentration $n_\romb$.  The average
counterion concentration is denoted by $\nb_\romc := |Q|/vV$.  In this
case the optimal screening constant can be expressed as
\begin{eqnarray}
  \kappab_\romopt^2
  & \stackrel{\text{(\ref{eq:kappab_v:v})}}{=} &
  \kappa^2 \cosh(v\psib_\romopt)
  \nonumber \\
  & \stackrel{\text{(\ref{eq:psibopt_determine_v:v})}}{=} &
  8 \pi \ell_\romB v^2 n_\romb \sqrt{1+(\nb_\romc/2n_\romb)^2}
  \nonumber \\
  & = &
  4 \pi \ell_\romB v^2 \sqrt{\nb_\romc^2+(2n_\romb)^2}.
  \label{eq:kappab_opt_v:v}
\end{eqnarray}
This expression differs from the screening constant of the salt
reservoir by also incorporating the contribution due to the
counterions.  It is illuminating to study its limiting behaviors at
high and low salt.  In the salt dominated regime, \ie\ $\nb_\romc \ll
2n_\romb$, the screening constant $\kappab_\romopt$ is practically
identical to the reservoir screening constant $\kappa$.  This is the
limit in which the Donnan effect is strongly diminished, and
Eqn.~(\ref{eq:psibopt_determine_v:v}) also shows that in this case
$\psib_\romopt\rightarrow 0$.  In the opposite limit of counterion
domination, $\nb_\romc \gg 2n_\romb$, the screening constant
$\kappab_\romopt$ is essentially determined by the average counterion
concentration, and it will be substantially larger than the screening
constant of the reservoir.  In this limit the Donnan effect is strong
and $\psib_\romopt$ is significantly different from the zero.  The
traditional linearization scheme $\psib=0$ will fail to adequately
describe this case, even though the ionic profiles can be very flat
and thus amenable to linearization.  In summary, we see that
linearizing the PB equation about $\psib=0$ invariably ignores the
counterion contribution to the screening, whereas the expansion about
the selfconsistently determined average potential includes the
counterions in a way that yields the correct behavior in the limits of
high and low salt.

The fact that a strong Donnan effect goes along with a value of
$\psib_\romopt$ substantially different from the reservoir potential
is also to be expected on more fundamental grounds: The difference in
microion concentrations across the membrane characteristic of the
Donnan equilibrium implies a concomitant jump of the electrostatic
potential, referred to as the ``Donnan-potential''.  Its magnitude can
likewise be used to measure the strength of the Donnan effect.  But a
large Donnan-potential also results in a large average potential.
This close connection between the Donnan-equilibrium and the optimal
linearization scheme lead the authors of Ref.~\cite{GrRo01} to suggest
the name ``Donnan-linearization'' \footnote{However, the optimal
linearization point $\psib_\romopt$ is not equal to the
Donnan-potential, because the boundary potential $\psi(R)$ is
different from the average potential $\langle\psi(\VECr)\rangle$.} for
the scheme from Eqn.~(\ref{eq:psibopt}).

The above findings can be succinctly reformulated in the following
way: By choosing the optimal linearization point $\psib_\romopt$ the
Donnan-equilibrium is automatically described correctly to lowest
order.  The solution of the linearized PB equation then provides the
higher order corrections.  We now show that \emph{every} other
linearization point gives a description that violates an important
zeroth-order inequality valid on the nonlinear PB level.  Let us first
specialize to a $v:v$ electrolyte.  If we denote by $\nb_\romc$ the
average density of counterions and by $\nb_\roms$ the average density
of salt molecules within the cell, we have within Poisson-Boltzmann
theory
\begin{eqnarray}
  (\nb_\romc+\nb_\roms)\,\nb_\roms & = &
  \langle n_+(\VECr) \rangle \, \langle n_-(\VECr) \rangle
  \; \stackrel{\text{(\ref{eq:Boltzmann})}}{=} \;
  n_\romb^2 \, \langle \rome^{-v\psi(\VECr)} \rangle \,
               \langle \rome^{ v\psi(\VECr)} \rangle
  \nonumber \\
  & \ge &
  n_\romb^2 \; \rome^{-v\langle\psi(\VECr)\rangle} \; \rome^{v\langle\psi(\VECr)\rangle}
  \; = \; n_\romb^2,
  \label{eq:ncns_ineq_PB}
\end{eqnarray}
where Jensen's inequality \footnote{If $f(x)$ is convex, its graph
lies above any of its tangents.  Constructing the tangent in $(\langle
x \rangle ; f(\langle x \rangle))$ shows that $f(x) \ge f(\langle
x\rangle) + (x-\langle x\rangle)f'(\langle x\rangle)$.  Averaging
yields Jensen's inequality $\langle f(x)\rangle \ge f(\langle
x\rangle)$.  If $f$ is concave, the inequality sign reverses.} has
been used.  Eqn.~(\ref{eq:ncns_ineq_PB}) states that the mean ionic
molarity \cite{Win95} within the suspension is greater than in the
reservoir.  On the other hand, within \emph{linearized}
Poisson-Boltzmann theory we find
\begin{eqnarray}
  (\nb_\romc+\nb_\roms)\,\nb_\roms
  & \stackrel{\text{(\ref{eq:lin_Boltzmann})}}{=} &
  n_\romb^2\,\big[1-v\langle\psi(\VECr)-\psib\rangle\big]
           \,\big[1+v\langle\psi(\VECr)-\psib\rangle\big]
  \nonumber \\
  & = &
  n_\romb^2\,\big[1-v^2\underbrace{\langle\psi(\VECr)-\psib\rangle^2}_{\ge 0}\big]
  \;\; \le \;\; n_\romb^2.
  \label{eq:ncns_ineq_LPB}
\end{eqnarray}
This is the inequality (\ref{eq:ncns_ineq_PB}) but \emph{with reversed
inequality sign}!  Observe that linearizing about $\psib_\romopt$ is
the \emph{only} scheme that does not violate (\ref{eq:ncns_ineq_PB}),
since Eqn.~(\ref{eq:psibopt}) renders (\ref{eq:ncns_ineq_LPB}) an
equality.  The importance of these two inequalities can be further
unveiled by solving them for the average salt concentration
$\nb_\roms$ inside the suspension, which gives the sequence
\begin{equation}
  \nb_\roms\big|_{\text{PB}}
  \; \ge \;
  \nb_\roms\big|_{\text{LPB, }\psib_\romopt}
  \; \ge \;
  \nb_\roms\big|_{\text{LPB}}
  \label{eq:ns_ineq}
\end{equation}
with
\begin{equation}
  2\,\nb_\roms\big|_{\text{LPB, }\psib_\romopt}
  \; = \;
  \sqrt{\nb_\romc^2 + (2\,n_\romb)^2} - \nb_\romc.
  \label{eq:ns_zeroth_order}
\end{equation}
In other words: Linearized PB theory gives a lower bound to the
average salt density within the cell, and $\psib_\romopt$ gives the
\emph{largest lower bound}.  Since underestimating $\nb_\roms$ means
overestimating the Donnan effect, the $\psib_\romopt$-scheme gives as
close a representation of the Donnan equilibrium as is possible in a
linear theory.  Observe also that the prediction
(\ref{eq:ns_zeroth_order}) for $\nb_\roms$ from Donnan linearization
is indeed the well known formula describing the Donnan effect after
neglecting activity coefficients (see for instance
Ref.~\cite{Ove56,Kek01}).  In that sense the set of inequalities
(\ref{eq:ns_ineq}) lies at the heart of the Donnan equilibrium and
distinguishes $\psib_\romopt$ as optimal.

It is possible -- although not straightforward -- to extend the above
considerations to general electrolyte compositions.  In that case one
has to look at more refined combinations of densities, also taking
into account to what percentage a particular species is represented in
the ionic mixture.  One possibility is the following: Define the
average
\begin{equation}
  \CALA \; := \;
  \sum_i p_i \log\frac{\langle n_i(\VECr)\rangle}{n_{i,\romb}}
  \quad\text{with}\quad
  p_i \; := \; \frac{n_{i,\romb}}{\sum_j n_{j,\romb}},
\end{equation}
which is inspired by a similar procedure employed when defining mean
activity coefficients \cite{Win95}.  Within PB theory we can readily
derive
\begin{eqnarray}
  \CALA & \stackrel{\text{PB}}{=} &
  \sum_i p_i \log\big\langle\rome^{-v_i\psi(\VECr)}\big\rangle
  \nonumber \\
  & \ge &
  - \sum_i p_i \, v_i \, \langle\psi(\VECr)\rangle
  \;\; = \;\;
  0,
  \label{eq:PB_ineq}
\end{eqnarray}
where Jensen's inequality has again been used.  The last step follows
from the charge neutrality of the salt reservoir.  However, within
linearized PB theory we get
\begin{eqnarray}
  \CALA & \stackrel{\text{LPB}}{=} &
  \sum_i p_i \log\big\langle \rome^{-v_i\psib}\big[1-v_i\big(\psi(\VECr)-\psib\big)\big]\big\rangle
  \nonumber \\
  & = &
  \sum_i p_i \log\big(1-v_i\big\langle\psi(\VECr)-\psib\big\rangle\big) 
  \nonumber \\
  & \le &
  -\sum_i p_i \, v_i \, \big\langle\psi(\VECr)-\psib\big\rangle 
  \;\; = \;\;
  0.
  \label{eq:LPB_ineq}
\end{eqnarray}
Here, the elementary inequality $\log(1+x)\le x$ has been used.  It is
straightforward to see that the relations (\ref{eq:PB_ineq}) and
(\ref{eq:LPB_ineq}) reduce to (\ref{eq:ncns_ineq_PB}) and
(\ref{eq:ncns_ineq_LPB}) for a $v:v$ electrolyte.  Nonlinear and
linearized PB theory again lead to conflicting inequalities in all but
one case: If $\psib_\romopt$ is chosen as the linearization point, the
PB inequality is not infringed.


\section{Explicit expressions and approximations for $d$-dimensional cell models}\label{sec:explicit}

In Sec.~\ref{sec:pressure} we gave expressions of the pressure in
terms of the charge- or potential-profiles that were based on the
derivative of the grand potential.  In this section we will
substantiate these results by inserting the actual solution of the
linearized PB equation for a $d$-dimensional cell model
\footnote{For an example how an analytical solution for a more complicated
cell (a finite cylinder) can be obtained see Ref.~\cite{TrHa97}.}.  We
will place particular emphasis on the formulas that emerge from using
the optimal linearization point $\psib_\romopt$ that we have discussed
in the last section.


\subsection{Analytical formulas for the pressure}

In the Appendix we outline how the linearized PB equation can be
solved for a general $d$-dimensional cell model (where $d=1$, $2$, and
$3$ corresponds to planar, cylindrical and spherical macroions,
respectively).  If we insert the final expression (\ref{eq:psi(R)})
into the pressure equation (\ref{eq:P_LPB_1_psi}), we get the explicit
formula
\begin{equation}
  \beta P_\romlin^{(1)} 
  \; = \;
  \sum_i \nb_i
  \, - \, \frac{1}{2} \Big[1-\gamma^2\Big(\frac{Q/V}{\sum_i v_i\,\nb_i}\Big)^2\Big]
  \frac{\big(\sum_i v_i \, \nb_i\big)^2}{\sum_i v_i^2 \, \nb_i},
  \label{eq:P1_explicit}
\end{equation}
where the variable $\gamma$ is defined as
\begin{equation}
  \gamma \; := \; \frac{1-\phi}{dD\sqrt{\phi}}
  \label{eq:gamma}
\end{equation}
and $D$ is given in Eqn.~(\ref{eq:determinant}).  In the case of
Donnan-linearization $-Q/V=\sum_i v_i\nb_i$ according to
Eqn.~(\ref{eq:psibopt_determine}), such that
Eqn.~(\ref{eq:P1_explicit}) further reduces to
\begin{equation}
  \beta P_\romlin^{(1)} 
  \;\; \stackrel{\psib_\romopt}{=} \;\;
  \sum_i \nb_i
  \, - \, \frac{1}{2} \big(1-\gamma^2\big)
  \frac{\big(\sum_i v_i \, \nb_i\big)^2}{\sum_i v_i^2 \, \nb_i}.
  \label{eq:P1_explicit_opt}
\end{equation}

The case of the alternative pressure definition (\ref{eq:P_psi_2}) is
less straightforward, since this depends explicitly on the whole
potential distribution $\psi(\VECr)$ and not just just on its boundary
value.  According to Eqn.~(\ref{eq:P_LPB_2}) the hard part is the
integral over $(\psi(\VECr)-\psib_\romopt)^2$, which is very unwieldy,
as it contains products of modified Bessel functions.  However, one
can always numerically integrate this expression or -- alternatively
-- numerically differentiate the grand potential
$\Omega_{\romlin,\romeq}$ from Eqn.~(\ref{eq:Omega_lin_eq}).

The explicit pressure formulas (\ref{eq:P1_explicit}) and
(\ref{eq:P1_explicit_opt}) are a further key result of this paper, and
in the remainder we will study some of their consequences.
Unfortunately, due to the algebraic complexity their properties cannot
be readily seen.  In the next subsection we will therefore spend some
time to study their analytical behavior in a few important limiting
cases.  Finally we will provide graphical illustrations for the
pressure formulas (\ref{eq:P1_explicit}) and
(\ref{eq:P1_explicit_opt}) as well as the pressure definition
(\ref{eq:P_psi_2}) for several exemplary cases in
Sec.~\ref{sec:examples}.


\subsection{Pressure bounds, limiting behavior and expansions}\label{ssec:bounds}

The pressure formula (\ref{eq:P_LPB_1_psi}) is quadratic in $\psi(R)$,
and it is easily checked that it takes its minimum value at the
inhomogeneity $\psi_\romi$ defined in Eqn.~(\ref{eq:psii}).  From
Eqn.~(\ref{eq:psi(R)}) and the definition of $\gamma$ in
Eqn.~(\ref{eq:gamma}) we see that this formally corresponds to
$\gamma=0$, and indeed the pressure expression (\ref{eq:P1_explicit})
attains its minimum value there.  For a symmetric $v:v$ electrolyte
this implies for the
\emph{excess} osmotic pressure
\begin{eqnarray}
  \frac{\beta \, \Delta P_{\romlin,\text{min}}^{(1)}}{2 n_\romb}
  & := &
  \frac{\beta(P_{\romlin,\text{min}}^{(1)}-P_{\text{res}})}{2 n_\romb}
  \nonumber \\
  & = &
  \cosh(v\psib) - \frac{1}{2}\sinh(v\psib)\tanh(v\psib) - 1
  \nonumber \\
  & = &
  \frac{\big[\cosh(v\psib)-1\big]^2}{2\cosh(v\psib)} \; \ge \; 0.
  \label{eq:P1_asympt}
\end{eqnarray}
Hence, in the symmetric case the pressure from linearized PB theory
defined via Eqn.~(\ref{eq:P_psi_1}) is nonnegative for every chosen
linearization point $\psib$.  However, this does not hold for general
electrolytes.  It may be verified that for the asymmetric case of a
$1:2$ electrolyte the excess pressure is negative if $\psib$ has the
same sign as the divalent ion species and $|\psib|\in[0;0.6309]$.  In
Donnan-linearization this comes down to the requirement that $Q$ has
the same sign as the divalent species and $0 \le |Q|
\le 1.596 n_\romb V$, where $n_\romb$ is the reservoir density of the
monovalent species.  Essentially, the pressure can become negative
when the counterion content within the cell is overwhelmed by the salt
ions.

Let us now restrict again to a symmetric $v:v$ electrolyte and define
the parameter
\begin{equation}
  \theta \; := \; \frac{|Q|/vV}{2n_\romb} \; = \; \frac{\nb_\romc}{2n_\romb}.
  \label{eq:theta}
\end{equation}
This ratio between average counterion concentration and salt reservoir
concentration measures whether we are in the salt dominated regime
($\theta$ small) or the counterion dominated regime ($\theta$ large).
Using $\theta$, Eqn.~(\ref{eq:kappab_opt_v:v}) can be rewritten as
$\kappab_\romopt^2=\kappa^2\sqrt{1+\theta^2}$.  The definition
(\ref{eq:theta}) shows that $\theta$ is small if either $V$ is large
(\ie, the volume fraction of colloids is low) or $n_\romb$ is large
(\ie, much salt has been added to the system).  In both cases
$\kappab_\romopt R$ becomes large and we may exploit the asymptotic
behavior of the modified Bessel functions \cite{AbSt70} to approximate
the quantity $D$ from Eqn.~(\ref{eq:determinant}) according to
\begin{eqnarray}
  D & \simeq &
  \romK_{d/2}(\kappab_\romopt r_0) \, \romI_{d/2}(\kappab_\romopt R)
  \nonumber \\
  & \simeq &
  \frac{\romK_{d/2}(\kappab_\romopt r_0)}{\sqrt{2\pi \kappab_\romopt R}}
  \, \rome^{\kappab_\romopt R}.
  \nonumber \\
  & \stackrel{\kappab_\romopt r_0\gg 1}{\simeq} &
  \frac{\rome^{\kappab_\romopt(R-r_0)}}{2\kappab_\romopt\sqrt{Rr_0}}
  \nonumber
\end{eqnarray}
Hence, $\gamma\propto1/D$ vanishes exponentially.  The pressure
$P_\romlin^{(1)}$ will thus approach its minimum value computed above.
Since in this limit $\psib_\romopt\rightarrow 0$, we may expand
Eqn.~(\ref{eq:P1_asympt}) for small $\psib_\romopt$ and obtain
\begin{equation}
  \frac{\beta \, \Delta P_{\romlin,\text{min}}^{(1)}}{2 n_\romb}
  \; \stackrel{\theta \ll 1}{\simeq} \;
  \frac{(v\psib_\romopt)^4}{8}
  \; \stackrel{(\ref{eq:psibopt_determine_v:v})}{\simeq} \;
  \frac{\theta^4}{8},
  \label{eq:P_asympt_0}
\end{equation}
showing that the excess osmotic pressure (measured in units of the
reservoir concentration) vanishes as the fourth power of $\theta$.  We
note in passing that the lowest order calculation based merely on the
average counterion concentration $\nb_\romc$ from
Eqn.~(\ref{eq:ns_zeroth_order}) (or, alternatively, the boundary
density from salt-free PB theory \cite{HaPo01}) gives instead the
asymptotic behavior $\theta^2/2$, \ie, only a quadratic dependence.

The full expression (\ref{eq:P1_explicit_opt}) contains a term which
appears to be of \emph{second} order in $\theta$, namely
$\theta^2\gamma^2/2$.  Since $\gamma$ vanishes exponentially, one
could be led to the incorrect conclusion (by ``prematurely''
terminating the expansion at this point) that the pressure in
linearized PB theory also vanishes exponentially \cite{RuBe81}.  While
it is in fact true that in the full \emph{nonlinear} theory the
pressure vanishes exponentially in the limit of salt-domination (a
simple plausibility argument for this can be found in
Ref.~\cite{DeHo01}), this does not hold for the linearized theory
\footnote{This conclusion remains valid if other pressure definitions are
used, for instance, the pressure equations (\ref{eq:P_PB_psi}) and
(\ref{eq:P_PB_n}) from nonlinear PB theory.}, where the exponential
decay of the term $\theta^2\gamma^2/2$ is masked by the quartic
asymptotic from Eqn.~(\ref{eq:P_asympt_0}).  In any case, one has to
be a little bit careful about the physical meaning of this limit.  It
indeed correctly describes a \emph{single} charged colloid immersed in
an electrolyte.  However, a dilute \emph{suspension} of many colloids
is not well described in this limit, since the lack of mutual
repulsions renders some basic assumptions about the cells
questionable.  Furthermore, one must keep in mind that our formulas
only give the osmotic pressure of the suspension due to the
\emph{microions}.  Even though this contribution most often dominates
just because there are many more microions than macroions, the
contribution of the latter has to become dominant once the microion
term vanishes exponentially.

The limit of large volume fraction, $\phi \rightarrow 1$, requires a
little more care.  Using $R=r_0\phi^{-1/d}$, an expansion of $D$
around $\phi=1$ gives to lowest order
\begin{eqnarray}
  D & \stackrel{\phi\rightarrow 1}{\simeq} &
  \frac{x_0}{2}\bigg\{\romK_{\nu+1}(x_0)\Big[\romI_\nu(x_0)+\romI_{\nu+2}(x_0)\Big]
  \nonumber \\
  & & \hspace*{-1em}
                    +\;\romI_{\nu+1}(x_0)\Big[\romK_\nu(x_0)+\romK_{\nu+2}(x_0)\Big]\bigg\}
  \times (\phi^{-1/d}-1),
  \nonumber
\end{eqnarray}
with $\nu=d/2-1$ and $x_0=\kappab_\romopt r_0=\kappa
r_0(1+\theta^2)^{1/4}$.  Since $\theta \propto 1/(1-\phi)$, both
$\theta$ and $x_0$ diverge as $\phi \rightarrow 1$.  $x_0/2$ times the
expression in curly brackets approaches $1-c/x_0^2$ with
$c=(d^2-1)(d^2+3)/64$ \cite{AbSt70}.  The term $\gamma$ therefore
behaves asymptotically like
\begin{equation}
  \gamma \; \stackrel{\phi\rightarrow 1}{\simeq} \;
  \frac{1-\phi}{d(1-c/x_0^2)(\phi^{-1/d}-1)\sqrt{\phi}}
  \; \simeq \; 1+\frac{c}{x_0^2}.
\end{equation}
This shows that $1-\gamma^2$ is of order $1/\theta$.  Using this, it
is now straightforward to show that to lowest order the high volume
fraction limit of Eqn.~(\ref{eq:P1_explicit}) is given by
\begin{equation}
  \beta P_\romlin^{(1)}
  \; \stackrel{\phi \rightarrow 1}{\simeq} \; \nb_\romc.
  \label{eq:P_asympt_1}
\end{equation}
This equation is simple to interpret: It states that the pressure is
given by the average counterion density.  This is reasonable, because
in this limit the ionic profiles become flat and thus the electrolyte
ideal.  Indeed, Eqn.~(\ref{eq:P_asympt_1}) merely states that the
osmotic coefficient goes to 1.  We hasten to add that
Eqn.~(\ref{eq:P_asympt_1}) only demonstrates the proper behavior
within the cell model.  Real suspensions crystallize at large enough
volume fraction, a transition which can only be described correctly
once one accounts for the ordered phase as well---see for instance
Refs.~\cite{HoAl83,ShAk87,ReBe97,KuDi97,RoHa97,RoDi99}.  But in order
to get the phase boundary right, one of course also needs a good
estimate of the grand potential in the fluid phase, so the correct
scaling of the optimal linearization at high volume fraction is
after all practically important.

Within Donnan-linearization positivity of the pressure does not
generally hold for the second pressure definition (\ref{eq:P_psi_2}),
not even for symmetric electrolytes.  For a $v:v$ electrolyte this can
be seen by differentiating Eqn.~(\ref{eq:psibopt_determine_v:v}) with
respect to $V$ and inserting into Eqn.~(\ref{eq:P_LPB_2}), whereby
one obtains
\begin{equation}
  \frac{\beta (P_\romlin^{(2)} - P_\romlin^{(1)})}{2 n_\romb}
  \; = \;
  -v^2\,\frac{\sinh^2v\psib_\romopt}{2\cosh\psib_\romopt}
  \big\langle\big(\psi(\VECr)-\psib_\romopt\big)^2\big\rangle,
  \label{eq:P2-P1}
\end{equation}
which is clearly negative.  In the limit of low volume fraction or
high salt $\psib_\romopt$ tends to zero, and expression
(\ref{eq:P2-P1}) scales like $(v\psib_\romopt)^2=\theta^2$ times
factors that do not tend to zero.  It hence does not vanish as quickly
as $P_\romlin^{(1)}\sim\theta^4$ and the pressure $P_\romlin^{(2)}$
must become negative in this limit.  We finally note that for
\emph{asymmetric} electrolytes $P_\romlin^{(2)}$ can also become
negative, but will not always be smaller than $P_\romlin^{(1)}$.

Let us close with a few remarks on the pressure in the traditional
linearization scheme $\psib=0$.  Since the linearization point is
volume-independent, the two pressure definitions (\ref{eq:P_psi_1})
and (\ref{eq:P_psi_2}) coincide.  If one calculates the excess osmotic
pressure by inserting $\psib=0$ into Eqn.~(\ref{eq:P_LPB_1_psi}), one
finds that the zeroth order is canceled by the reservoir pressure,
while the first order drops out due to reservoir electroneutrality,
giving
\begin{eqnarray}
  \beta\,\Delta P_\romlin
  & \stackrel{\psib=0}{=} &
  \frac{1}{2}\psi(R)^2\sum_i v_i^2 n_{i,\romb}
  \nonumber \\
  & \stackrel{(\ref{eq:psi(R)})}{=} &
  \frac{(Q/V)^2}{2\sum_i v_i^2 n_{i,\romb}}\,\gamma^2.
  \label{eq:P_psib=0}
\end{eqnarray}
In the special case of a $v:v$ electrolyte this simplifies further to
$\beta\Delta P_\romlin/2n_\romb = \theta^2\gamma^2/2$ (recognize the
``second order'' term from above!).  Eqn.~(\ref{eq:P_psib=0}) shows
that this pressure is always positive.  Moreover, in the limit of low
$\phi$ or large salt it vanishes exponentially as
$\gamma^2\sim\exp\{-2\kappa R\}$ -- just as in full PB theory.  Even
though $\psib_\romopt$ also approaches zero in this limit, the
corresponding pressure asymptotics is dominated by the way in which it
does so, yielding a power law as discussed above.  In the opposite
limit of large $\phi$ we have seen above that $\gamma\rightarrow 1$,
hence for the $v:v$ case the pressure using $\psib=0$--linearization
asymptotically behaves like $\beta \Delta P_\romlin \sim
\nb_\romc^2/4n_\romb$, \ie, it diverges quadratically in $\nb_\romc$
and not linearly as it should.  We finally note that a quadratic
expression like (\ref{eq:P_psib=0}) has recently been rederived in
Ref.~\cite{ZhCz01} by using a perturbative expansion of the PB
equation and restricting to lowest order.  However, in this approach
it is difficult to see that this is in fact thermodynamically
consistent.


\section{Exemplary comparison with nonlinear PB theory}\label{sec:examples}

In this section we provide a comparison of the above pressure formulas
with the results from nonlinear PB theory \footnote{The nonlinear PB
equation for the cell model has been solved by discretizing the radial
coordinate, starting with a trial density distribution and alternately
($i$) computing the potential from the charge distribution via
Poisson's equation and ($ii$) determining the ion distributions from
the Boltzmann relation (\ref{eq:Boltzmann}) until selfconsistency has
been achieved.  Due to the Donnan equilibrium this has to be done
under the additional constraint that the chemical potential of the
small ions has the same value as in the salt reservoir.  For a $v:v$
electrolyte this for instance implies that $n_+(\VECr)n_-(\VECr) =
n_\romb^2$.  See also Ref.~\cite{TaLe98}.}, which is intended to
clarify and illustrate the findings from the last sections.  Apart
from the pressure, we will also calculate the \emph{compressibility},
which is defined as $\CALK = \frac{1}{V}(\partial V/\partial \Delta
P)$.  For a graphical representation it is however more convenient to
plot the reduced inverse compressibility $\partial \beta\Delta
P/\partial \nb_\romc = 1/\nb_\romc k_\romB T \CALK$, where the average
density of counterions is again defined by $\nb_\romc=|Q|/vV$.

In order to underline the generality of the formulas derived above, we
will use the following three different situations: ($i$) spherical
colloids in a 1:1 electrolyte, ($ii$) cylindrical colloids in a 1:1
electrolyte, and ($iii$) spherical colloids in a 1:2 electrolyte.

\vspace*{1em}


\begin{figure*}
  \includegraphics[scale=1.05]{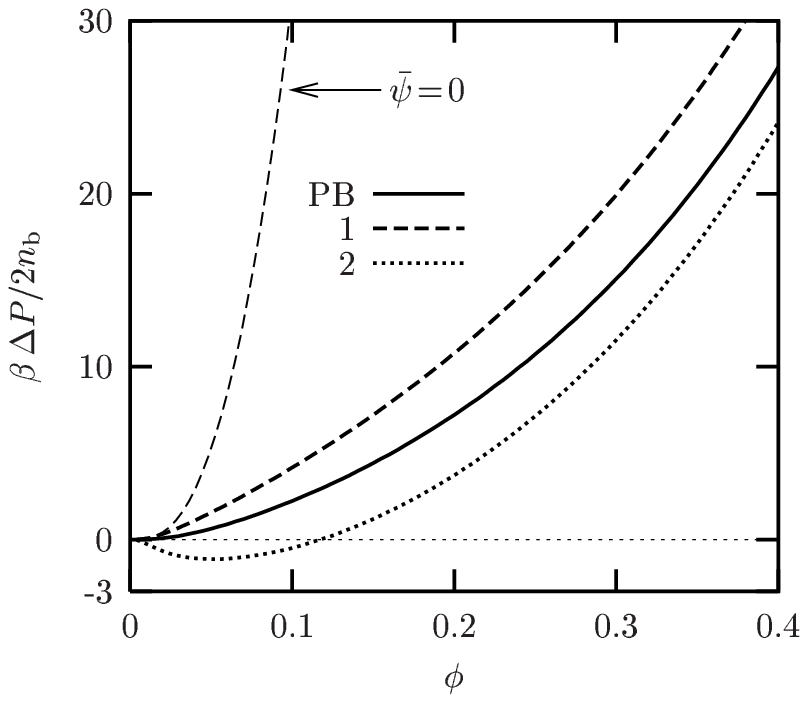}\hfill
  \includegraphics[scale=1.05]{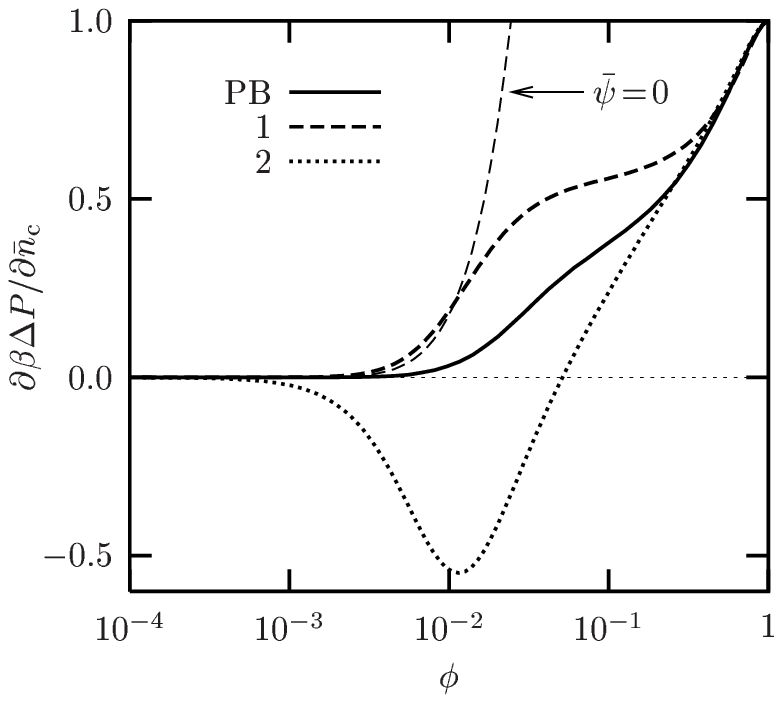}
  \caption{Pressure (left) and inverse reduced compressibility (right)
  as a function of volume fraction $\phi$ for a solution of
  \emph{spherical} colloids having a bare charge $Q=3500\,e$ and a
  radius $r_0=133\,\text{nm}$.  The system is in contact with a dilute
  1:1 electrolyte of concentration $3.6\,\mu\text{M}$ and the Bjerrum
  length is $\ell_\romB=\,0.714\,\text{nm}$, corresponding to water at
  room temperature.  The solid curve is the result from nonlinear PB
  theory, the bold dashed curve is the first pressure definition
  (\ref{eq:P_LPB_1_psi}) combined with the optimal linearization from
  Sec.~\ref{sec:psibb} (leading to the explicit expression
  (\ref{eq:P1_explicit_opt})) while the dotted curve combines this
  scheme with the second pressure definition (\ref{eq:P_LPB_2}).  The
  fine dashed curve uses $\psib=0$, in which case both pressure
  definitions coincide, leading to the explicit formula
  (\ref{eq:P_psib=0}).}
  \label{FIG:3}
\end{figure*}

The first system we study consists of spherical colloids of charge
$Q=3500\,e$ and radius $r_0=133\,\text{nm}$ immersed in an aqueous
solution (\ie, $\ell_\romB=\,0.714\,\text{nm}$) which is dialyzed
against a 1:1 electrolyte of rather low molarity $3.6\,\mu\text{M}$.
Motivated by the prediction of a gas-liquid phase separation at these
parameters \cite{RoDi99}, the authors of Ref.~\cite{GrRo01} used them
as an illustration which pushes linearized theory to the limits of
validity, making the way in which it deviates from PB theory
particularly visible.  Lowering the colloidal charge entails a
successively better agreement between linear and nonlinear theory, as
can also be seen in Ref.~\cite{GrRo01}.

Fig.~\ref{FIG:3} shows the pressure (left) and the inverse
reduced compressibility (right) as a function of volume fraction
$\phi$.  The solid line corresponds to the solution of the nonlinear
PB equation.  The pressure is given by Eqn.~(\ref{eq:P_PB_n}) and the
excess pressure is always positive (see Eqn.(\ref{eq:PB_Pge0})).  The
same holds for the compressibility \footnote{While $P\ge 0$ for the PB
cell model follows easily (see Eqn.~(\ref{eq:PB_Pge0})), the
corresponding inequality $\CALK \ge 0$ is much less trivial, and to
the author's knowledge no formal proof has yet been given.}.  The
dashed and dotted curves correspond to the solutions from linearized
PB theory.  For $\psib=0$ both pressure definitions (\ref{eq:P_psi_1})
and (\ref{eq:P_psi_2}) coincide, but in this case the predictions are
clearly off from the PB result except at very low volume fraction.  In
contrast, using the optimal linearization point $\psib_\romopt$ from
Sec.~\ref{ssec:psiopt} brings about the correct behavior at high
volume fraction, \ie, in the regime where the cell model is
particularly appropriate and linearization should indeed work because
of relatively flat ionic profiles.

The pressure definition (\ref{eq:P_psi_2}) -- the \emph{total}
derivative of the grand potential $\Omega(V, T, \ldots,
\psib_\romopt(V))$ with respect to volume -- corresponds to the
one employed in Ref.~\cite{GrRo01}.  It can be seen to lead to
negative pressures and compressibilities at moderately low volume
fractions, which would imply a segregation into a dense and a dilute
colloidal phase.  Such a phase transition has also been claimed in
other recent theoretical works \cite{RoHa97,RoDi99,War00} which
likewise are essentially based on linearized PB theory.  However, the
similarity of the resulting phase diagrams, the appearance of
essentially the same ``volume terms'' in the grand potential (which
here are responsible for the effect), and the fact that full PB theory
does not show any phase transition, lead the authors of
Ref.~\cite{GrRo01} to question these claims of a gas-liquid
coexistence in such systems as a spurious side effect of
linearization.  In a similar spirit, Diehl \etal\ \cite{DiBa01} show
(within a generalized Debye-H\"uckel-Bjerrum approach
\cite{LeBa98,TaLe98_2}) that explicitly re-incorporating the effects
of counterion condensation, which are neglected when doing the
linearization, removes (or at least strongly suppresses) the phase
transition otherwise clearly visible.  We remark that Fig.~1 of
Ref.~\cite{DiBa01}, showing the excess osmotic pressure as a function
of volume fraction for a salt-free suspension of spherical colloids
with varying bare charge, can be reproduced almost quantitatively
within the cell model by using Donnan-linearization and employing the
pressure definition (\ref{eq:P_psi_2}).  While these warnings about
the dangers of linearization are certainly well made, we want to point
out that things are in fact even a little more subtle: Not even
linearized theory needs to show the phase transition.  Whether or not
it does so depends crucially on the pressure definition as well as the
linearization point.  For $\psib=0$ there is no phase transition.  For
$\psib=\psib_\romopt$ there is a phase transition only if pressure
definition (\ref{eq:P_psi_2}) is used.  The definition
(\ref{eq:P_psi_1}) -- which is based on the \emph{partial} derivative
of the grand potential with respect to volume -- yields a positive
compressibility.

Let us repeat that this pressure definition (\ref{eq:P_psi_1}) could
only be written down once the linearization point has been recognized
as an independent variable.  And since it is the volume-dependence of
$\psib_\romopt$ that renders $P_\romlin^{(2)}$ negative, it must also
be related to the ``volume'' terms discussed in
Refs.~\cite{RoHa97,RoDi99,War00,GrRo01} (see also Ref.~\cite{Cha01}).
Our formulation of the problem hence illustrates that the way in which
these terms drive a phase transition is closely related to the choice
which variables one intends to keep constant when differentiating the
thermodynamic potential.  Within the cell model it is obvious that PB
theory only yields positive pressures and which pressure definition is
hence preferrable, but when relaxing the constraints of a cell model
or mean field PB theory the nature of effective interactions is much
less obvious.  See for instance Ref.~\cite{Bel00} for a recent
critical evaluation of the theoretically as well as experimentally
subtle issue of phase separation in suspensions of spherical colloids.


\vspace*{0.5em}

\begin{figure*}
  \includegraphics[scale=1.05]{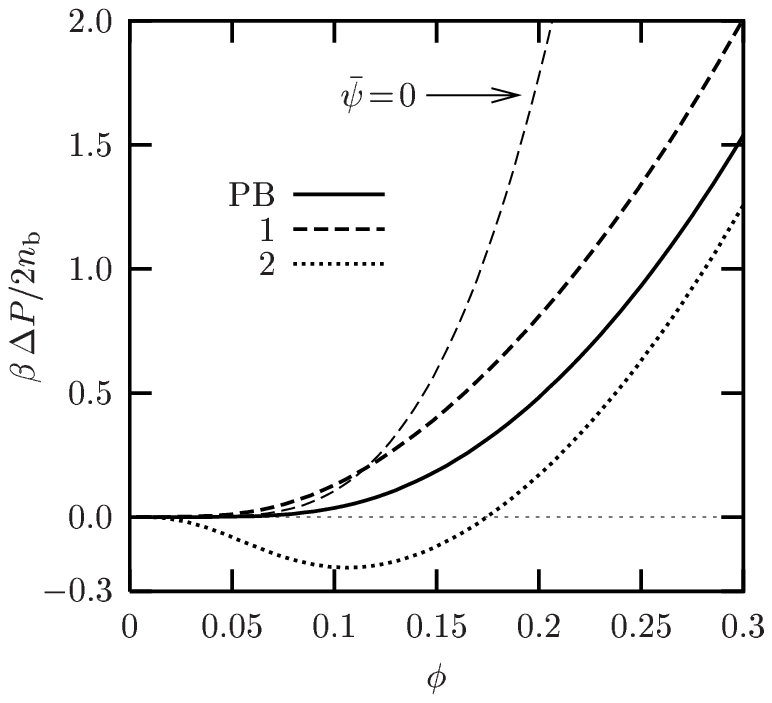}\hfill
  \includegraphics[scale=1.05]{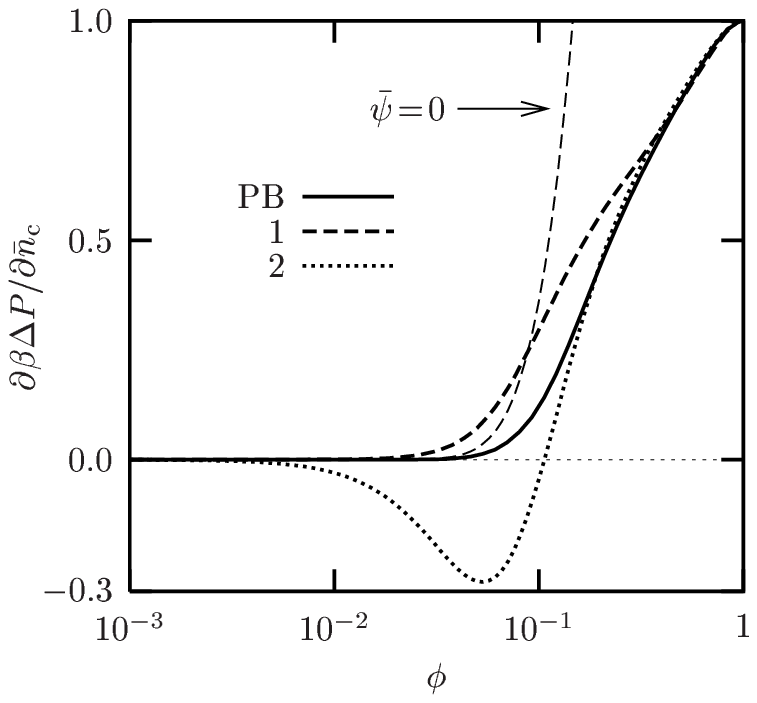}
  \caption{Same plots as in Fig.~\ref{FIG:3}, but for
  \emph{cylindrical} macroions having a charge parameter $\xi=4.2$
  (\ie, 4.2 charges along the axis per Bjerrum length) and a radius
  $r_0=1.2\,\text{nm}$ (those values correspond roughly to DNA).  The
  system is in contact with a dilute 1:1 electrolyte of physiological
  salt concentration $100\,\text{mM}$. The line styles
  are the same as in Fig.~\ref{FIG:3}.}
  \label{FIG:4}
\end{figure*}

In our next example we turn to a physical system which differs in the
geometry of the colloids and which is (quite literally) of vital
importance.  Fig.~\ref{FIG:4} shows pressure and
compressibility of an aqueous solution of cylindrical colloids, \ie,
charged rods, which have a radius of $r_0=1.2\,\text{nm}$ and a line
charge density of 1 charge per $1.7\,\text{\AA}$ (or 4.2 charges per
Bjerrum length) dialyzed against an 1:1 electrolyte of concentration
$100\,\text{mM}$.  The colloid is much stronger charged than in the
spherical situation above, in the sense that the surface charge density
is a factor 50 larger.  On the other hand, the salt concentration is
also much larger (with the reservoir screening length being only 0.6\%
of the screening length in the above low-salt case) which keeps the
potentials low.  The above choice of values corresponds to DNA in a
physiological salt environment.  As Fig.~\ref{FIG:4} shows,
nonlinear PB theory again gives positive pressure and compressibility
for all volume fractions $\phi$, and the large-$\phi$ behavior is
captured correctly if Donnan linearization is employed, while the
choice $\psib=0$ fails there.  However, using $\psib_\romopt$ and the
definition (\ref{eq:P_psi_2}) results in a pressure that is negative
for $\phi \lesssim 17\%$ and a negative compressibility for $\phi
\lesssim 11\%$.  If this were true, all DNA in animal cells would tend
to aggregate and phase separate!  But again, this failure is not an
inevitable artifact of linearization, since the pressure
$P_\romlin^{(1)}$ is perfectly positive and gives rise to positive
compressibilities.  Incidentally, DNA \emph{can} be condensed, but
this requires multivalent ions \cite{Blo91} and is known to be a
correlation effect which is missing in PB theory, see for instance
\cite{GuNi86,LyNo97,GrMa97}.

Poisson-Boltzmann theory and the linearization scheme employing $\psib
= 0$ predict an exponentially vanishing pressure as $\phi \rightarrow
0$.  However, as we have remarked above the pressure due to the
\emph{macroions} has to become significant at some point (see
Ref.~\cite{RaCo00} for measurements on DNA in low salt solutions which
appear to support this view).  For linear polyelectrolytes this is
even more relevant than for spherical colloids, since the
conformational degrees of freedom and the degree of entanglement
becomes important \cite{DoCo95}.


\vspace*{0.5em}

\begin{figure*}
  \includegraphics[scale=1.05]{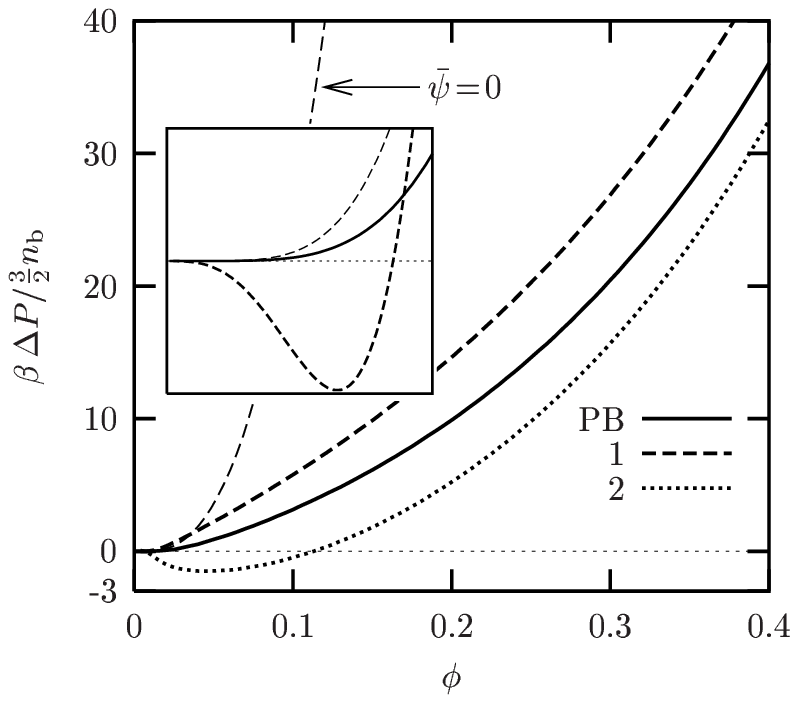}\hfill
  \includegraphics[scale=1.05]{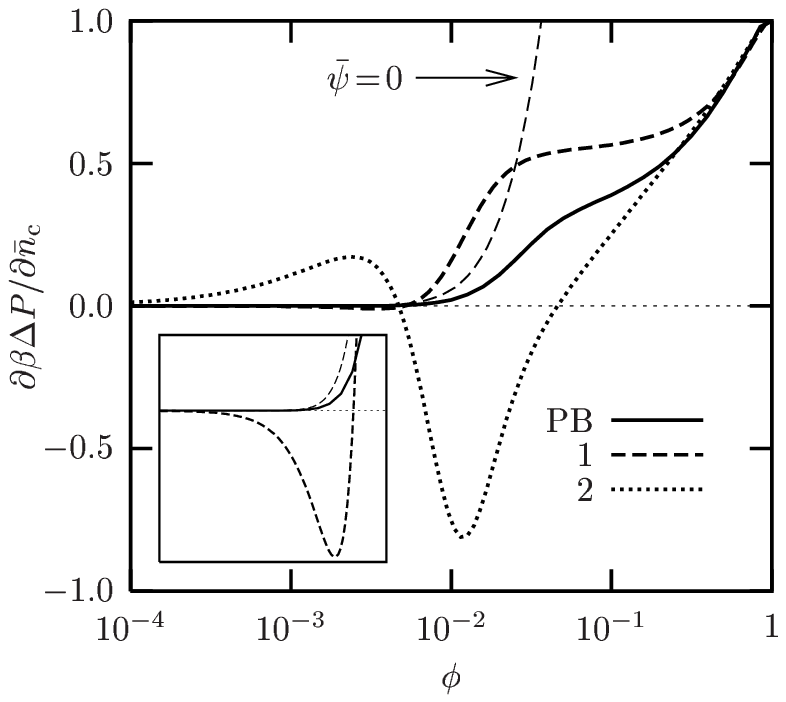}
  \caption{Same plots as in Fig.~\ref{FIG:3}, but for a 1:2
  electrolyte.  The colloid charge has the same sign as the divalent
  ions, the reservoir density of the monovalent ions is $3.6\,\mu\text{M}$.
  The inset in the left panel magnifies the regions
  $[0;0.008]\times[-0.003;0.003]$, while the inset on the right
  magnifies $[10^{-4};10^{-2}]\times[-0.01;0.005]$, showing that in
  this example the first pressure definition (\ref{eq:P_psi_1}) also
  gives negative pressures and compressibilities, while the PB result
  does not.}  \label{FIG:5}
\end{figure*}

In our final example we study an asymmetric electrolyte.  We go back
to the system of spherical colloids studied above and replace the ions
of one sign by divalent ones, \ie, we assume a 1:2 electrolyte of
concentration $3.6\,\mu\text{M}$, which in particular implies that the
divalent species occurs with a concentration of $1.8\,\mu\text{M}$ in
the reservoir.  According to the discussion in Sec.~\ref{ssec:bounds}
the case in which the colloid has the same sign of charge as the
divalent ions is particularly interesting, since then positivity of
the pressure cannot be guaranteed even for the pressure definition
(\ref{eq:P_psi_1}).  For this case Fig.~\ref{FIG:5}
shows again pressure and compressibility as a function of volume
fraction.  PB theory is once more found to give positive results, and
the same remarks as above about the poor high-$\phi$ behavior of the
linearization point $\psib=0$ compared to $\psib=\psib_\romopt$ apply
also here.

However, there is an (expected) difference concerning the behavior of
$P_\romlin^{(1)}$, which becomes (very slightly) negative for $\phi
\lesssim 0.68\%$ and gives rise to negative compressibility for $\phi
\lesssim 0.51\%$, as is illustrated in the insets.
In Sec.~\ref{ssec:bounds} we have seen that the pressure \emph{can}
become negative if $|Q| \le 1.596 n_\romb V$, which in this case
implies $\phi \lesssim 0.96\%$ -- and indeed, lowering $\phi$ by
further 30\% brings about the negative pressure.  We mention in
passing that with a colloid having the opposite sign of charge as the
divalent species this does not occur.


\section{Conclusion and summary}

The main motivation of the present paper has been the following
question: How can the membrane pressure due to a Donnan-equilibrium
best be described within linearized PB theory?  This first required a
clarification of ($i$) how one computes the pressure and ($ii$) what
one means by ``best''.  Often the pressure has been computed (without
much justification) by using the predictions of the linearized PB
equation in formulas of the nonlinear theory (or expansions thereof);
moreover, this recourse to the full theory is not unique and hence
prone to inconsistencies.  These problems are avoided by starting from
the appropriate thermodynamic potential functional of linear theory,
which however still leaves two alternative definitions for the
pressure that differ in their treatment of a possible volume
dependence of the linearization point.  Neither definition predicts
the pressure to be proportional to the density of microions at the
outer cell boundary, \ie, the well known result from nonlinear PB
theory does not apply on the linearized level.  All this can be
discussed without fixing the expansion point, but optimality
considerations nevertheless suggest a specific choice for it.
Explicit formulas can be obtained, since the linearized PB equation
can be solved analytically for general cell models.

Let us thus sumarize the key findings of this work:

(a) Linearized PB theory can be based on a density functional which is
the quadratic expansion of the well known functional of nonlinear
PB theory.  The choice of the expansion point distinguishes different
linearization schemes.  The equilibrium value of this functional is
the thermodynamic potential, in our case the grand potential.

(b) The pressure is given by the volume derivative of this
thermodynamic potential.  There is a choice as to whether or not one
would like to keep the linearization point fixed, resulting in two
different pressure definitions both based on the thermodynamic
potential.

(c) The boundary density rule (\ref{eq:P_PB_n}) from nonlinear PB
theory is replaced by the quadratic expansion of the nonlinear formula
in the boundary potential.

(d) The range of validity of linearization depends on the expansion
point.  We proved that linearization about the selfconsistently
determined average potential is optimal---in the sense that it
automatically describes the Donnan effect correctly in lowest order,
leaving the linearized PB equation to incorporate higher order
corrections.  Furthermore, we showed that all other linearization
schemes violate an important inequality from nonlinear PB theory
related directly to the Donnan effect.

(e) The linearized PB equation can be solved exactly for symmetric
cell models of arbitrary dimension, salt reservoir composition and
linearization point.  We used this solution to derive explicit
formulas for the pressure and have discussed their analytical
properties in detail.

(f) A comparison with the results from nonlinear PB theory has been
performed, showing that the validity of linearization depends strongly
on the linearization point.  While the traditional linearization
scheme completely fails to describe the important limit of large
volume fraction, the optimal (Donnan-) linearization becomes
asymptotically correct there.

(g) For pressure definition (\ref{eq:P_psi_2}), \ie\ the \emph{total}
volume derivative of the grand potential, negative pressures and
compressibilities can occur.  These findings have previously been
taken as indications of a gas-liquid phase separation in suspensions
of spherical colloids.  In the present work we showed that the same
theoretical reasoning would predict DNA to phase separate under
physiological conditions at all relevant concentrations.  We consider
this as a further striking argument against the genuineness of the
transition.

(h) In nonlinear PB theory the pressure is always positive.  Whether
or not this still holds on the linearized level depends crucially on
the precise definition of the pressure itself.  It is for this reason
that we advocate defining the pressure via the \emph{partial}
derivative of the grand potential with respect to the volume (\ie,
keeping in particular the linearization point fixed), since the
pressure is then always positive for symmetric electrolytes---in
agreement with nonlinear PB theory.


\begin{acknowledgments}
MD would like to thank M.\ Tamashiro, I.\ Borukhov, and H.\
Wennerstr\"om for stimulating discussions or clarifying comments
during the development of this work.  He also gratefully acknowledges
financial support by the German Science Foundation (DFG) under grant
De775/1-1.  HHvG gratefully acknowledges intensive discussions with
R.\ Klein and R.\ van Roij.
\end{acknowledgments}


\appendix

\section{Solution of the linear PB equation for a cell model in $d$ dimensions}

Consider a cell model in $d$ dimensions with generalized radial
coordinate $r$. $d=1$, $2$, and $3$ corresponds to planar, cylindrical
and spherical macroions, respectively.  After the transformation $x =
\kappab r$ and $\psi(r) = \tilde\psi(\kappab r) = \tilde\psi(x) $ the
linearized PB equation (\ref{eq:LPB}) reads
\begin{equation}
  \tilde\psi''(x) + \frac{d-1}{x}\,\tilde\psi'(x)
  \; = \;
  \tilde\psi(x) - \psi_\romi,
  \label{eq:PBlinDE}
\end{equation}
where the prime indicates differentiation with respect to $x$.
Henceforth we will not bother with the difference between $\tilde\psi$
and $\psi$ and omit the tilde.  An obvious particular solution of
Eqn.~(\ref{eq:PBlinDE}) is $\psi(x)=\psi_\romi$, while the homogeneous
equation is solved by the ansatz $\psi(x) = w(x)/x^\nu$ with $\nu =
d/2-1$, provided $w(x)$ solves the Bessel equation $x^2\,w''(x) +
x\,w'(x) - (x^2+\nu^2)\,w(x) = 0$.  Two linearly independent solutions
are the modified Bessel functions $\romK_\nu$ and $\romI_\nu$
\cite{AbSt70}.  The general solution of the differential equation
(\ref{eq:PBlinDE}) can therefore be written as
\begin{equation}
  \psi(x) =
  \psi_\romi
  + C_\romK\,\frac{\romK_\nu(x)}{x^\nu}
  + C_\romI\,\frac{\romI_\nu(x)}{x^\nu},
  \label{eq:PBlinDGLallgLoesung}
\end{equation}
where $C_\romK$ and $C_\romI$ are constants to be determined by the
following boundary conditions.  Gauss' law relates the radial
derivative of the electrostatic potential at $r_0=x_0/\kappab$ to the
surface charge density $e\sigma$ of the macroion.  If we define the
latter to be the total charge of the macroion divided by its surface
area, we have $\psi'(x_0) = -4\pi\ell_\romB\sigma/\kappab$.  Since the
whole cell is neutral, we also have $\psi'(X) = 0$ at the outer cell
radius $R=X/\kappab$.  The integration constants now follow from
inserting these boundary conditions into the general solution.  After
a little algebra we then find the potential
\begin{eqnarray}
  \psi(x) & = &
  \psi_\romi +
  \frac{4\pi\ell_\romB\sigma}{\kappab D}
  \left(\frac{x_0}{x}\right)^\nu
  \nonumber \\
  & & \times
  \, \Big[\romI_{\nu+1}(X)\romK_\nu(x) + \romK_{\nu+1}(X)\romI_\nu(x)\Big],
  \label{eq:FVL}
\end{eqnarray}
where the determinant $D$ is defined as
\begin{equation}
  D \; = \; \romK_{\nu+1}(x_0)\romI_{\nu+1}(X) - \romK_{\nu+1}(X)\romI_{\nu+1}(x_0).
  \label{eq:determinant} 
\end{equation}
We have $D>0$ since $X>x_0$.

At the outer boundary the potential (\ref{eq:FVL}) can be further
simplified.  The term in angular brackets reduces to $1/X$, and using
the volume fraction $\phi$ the surface charge density can be rewritten
in terms of the colloid charge as $\sigma/r_0 = (Q/V)(1-\phi)/d\phi$.
Inserting the definitions (\ref{eq:kappab}) for $\kappab$ we arrive at
\begin{equation}
  \psi(R) \; = \;
  \psi_\romi + \frac{Q/V}{\sum_i v_i^2 \, \nb_i}\frac{1-\phi}{dD\sqrt{\phi}}.
  \label{eq:psi(R)}
\end{equation}
Equation (\ref{eq:psi(R)}) gives the boundary potential in linearized
PB theory for a general $d$-dimensional cell model with arbitrary
linearization point $\psib$ and electrolyte composition.



\end{document}